\documentclass[useAMS,usenatbib]{mn2e}

\title[Radio emission threshold]{The threshold for pulsar radio emission is determined by the Goldreich-Julian charge density}
\author[P. B. Jones]{P. B. Jones\thanks{E-mail:
peter.jones@physics.ox.ac.uk}  \\
University of Oxford, Department of Physics, Denys Wilkinson Building,\\
Keble Road, Oxford OX1 3RH, U.K.}

\begin{document}

\date{}

\pagerange{\pageref{firstpage}--\pageref{lastpage}}
\pubyear{}

\maketitle

\label{firstpage}

\begin{abstract}
A recent phenomenological study of radio emission from normal and millisecond pulsars by Karastergiou et al has lead these authors to state that they are unable to exclude a common physics process as the source although the rotation periods and magnetic fields of these two classes are very different.  This has bearing on the nature of that source and it is the purpose of the present Letter to explore this problem further, specifically, for the ion-proton model and for all those models that assume electron-positron pair creation above the polar cap.  The ion-proton model satisfies this commonality and we briefly mention some consequences of this finding.
\end{abstract}


\section{Introduction}

Karastergiou et al (2024) have made a careful phenomenological study of pulsar radio emission, comparing normal pulsars with millisecond pulsars (MSP). Profile widths $W_{10}$ have a common power-law dependence on the rotation period $P$, also the distributions of spectral indices are very closely similar.  These authors conclude, tentatively, that the data support common emission physics between normal and millisecond pulsars (MSP) although there are orders of magnitude differences between them in $P$ and $\dot{P}$. Their paper stimulated the work of the present Letter though with less emphasis on phenomenology and with the introduction of specific model physical mechanisms. Any such model must have some parameter(s) in its description but the basic neutron-star parameters from which they can be constructed are, at present, limited to those listed in the ATNF Catalogue (Version 2.5.1; Manchester et al 2005) specifically, the rotation period and the polar-cap magnetic flux density $B_{12}$, in units of $10^{12}$ Gauss derived from the observed period time derivative. The commonality found by Karastergiou et al requires common values of these constructed parameter(s) at the threshold of radio emission for both normal pulsars and MSP.

Apart from papers based on electron-positron pair production, the ion-proton model appears to be the only published mechanism (see Jones 2023). This requires some explanation.  The amplitude growth exponent of the Langmuir mode concerned is given by,
\begin{eqnarray}
\Lambda = w\left(\frac{4\pi(1 - \alpha_{p})Z_{\infty}eR^{2}B}{PAm_{p}c^{3}}\right)^{1/2}\int^{\infty}_{1}
d\eta\left(\frac{1}{\eta\gamma_{z}(\eta}\right)^{3/2},
\end{eqnarray}
in which the dimensionless prefactor $w$ is found from the roots of the plasma dispersion relation and $w = 0.2$ is a typical value. The ion mass and charge are $A,Z_{\infty}$ and $m_{p}$ is the proton mass.  The neutron star radius is $R$ and $B$ its polar-cap surface field. The radius from the neutron-star centre is $R\eta$, $\gamma_{z}$ the ion Lorentz factor, and $\alpha_{p}$ the fraction of the polar-cap current carried by protons.  The value of the integral in equation (1) depends on the degree of screening of the accelerating electric field and, being self-adjusting, is difficult to estimate.

Thresholds for pair production are totally dependent on unknown flux-line curvatures above the polar cap and on primary electron Lorentz factors.  In contrast, the baryons in the ion-proton model have relativistic, though small, Lorentz factors that are necessary for adequate amplitude growth-rate exponents of the longitudinal Langmuir mode that is the source of the radiation.  Equation (1) shows that in terms of common neutron-star parameters this exponent scales as $(B_{12}P^{-1})^{1/2}$ which is a suitable  parameter for this model. In the case of pairs, the maximum potential difference in a vacuum region above the polar cap is approximately $8\times 10^{3}B_{12}P^{-2}$ GeV, but account must also be taken of the single-photon pair-creation threshold for which we adopt the approximate practical expression $k_{\perp} = 9mc(B_{12})^{-1}$ for a photon momentum $k$.

An approach to the radius of curvature problem that has been adopted previously is that a curvature photon emitted in the open sector of the magnetosphere above the polar cap is required to create a pair in that sector but not in the closed sector beyond (see Jones 2021). The simple model adopted is that of assuming that the open-sector field remains constant at the surface value above the polar cap, that the open sector and all flux lines within it have a constant radius of curvature and that photons are emitted tangentially to flux lines. From the geometry of chords within a circle, for a photon of momentum $k$, this can be expressed approximately as $k_{\perp} > k(2u_{0}/\rho)^{1/2}$ at the point of pair creation, where $u_{0}$ is the polar-cap radius and $\rho$ is the flux-line radius of curvature. Allowance for the decrease in field with altitude in a more refined model would decrease the probability of pair creation. The value of the electron Lorentz factor $\gamma$ is assumed to be given by the maximum vacuum potential difference as in the work of Philippov, Timokhin and Spitkovsky (2020). For the pulsars listed in Table 2, imposing the geometrical condition above to solve for the threshold value of the radius of curvature above which pair creation is not possible, gives radii of the order of $10^{6}$ cm. But this in itself is indefinite and not useful. So the chosen parameter for pair creation is $B_{12}P^{-2}$ which scales the energy of the primary accelerated electron and for curvature radiation its cube scales the photon energy. Its choice as basic parameter for pair creation is not ideal but is possibly the optimum.  This procedure for dealing with the radius of curvature problem has been followed previously for the MSP J0030+0451 (Jones 2021) with the result that pair production is only very marginally possible in that pulsar.  An observational test for the existence of the photon-pair cascade was proposed in that Letter consisting of a search for GeV photons in approximate phase with the radio profile, but there is no indication that it has been performed or satisfied.

\section{Comparison of the models}

 This Letter seeks to compare values of the two parameters described in the previous Section for pulsars at the threshold of observability, and it is necessary to say how this has been defined. The normal pulsars and SMP have been drawn from the complete ATNF Catalogue by setting conditions on rotation period and on $BP^{-1}$. Table 1 gives the threshold parameters for all those pulsars with $P<0.2$s and $BP^{-1} < 2.5\times 10^{10}$ G$s^{-1}$. The latter constraint has been obtained by adjusting its value upwards until a suitable number of pulsars are allowed. Table 2 pulsars are similarly chosen with $P > 0.2$s and $BP^{-1} < 1.0\times 10^{11}$ G$s^{-1}$.  All except J0514-4002A and J2034+3632 have catalogued values of $W_{10}$ and/or $S_{1400}$ which supports our assumption that they are neutron stars.Apart from these threshold cases, the remaining pulsars in the ATNF catalogue will naturally  have wide distributions of parameter values, for example, the parameters for young normal pulsars are not to be compared with MSP which have a different history. This the reason why the present Letter compares threshold values.  Each Table starts with the J-name and gives the values of the constructed parameters $B_{12}P^{-2}$ and $(B_{12}P^{-1})^{1/2}$ that are observed.  The important point in comparison of the two Tables is not that each has a narrow distribution of $(B_{12}P^{-1})^{1/2}$, which is to be expected because the pulsars are within small but well separated areas of the $P - \dot{P}$ plane, but that the average values for the two Tables, being 0.14 and 0.28 respectively, differ only by a modest factor, which could be explained by systematic differences between the two classes of pulsar in factors such as $\alpha_{p}$ and $Z_{\infty}$ in equation (1), bearing in mind that they have different histories.   This is in contrast with values for $B_{12}P^{-2}$ which differ typically by one or two orders of magnitude between normal pulsars and MSP. The averages are 5.0 and 0.18 respectively for Tables 1 and 2. Thus the conclusion of this Letter is that $(B_{12}P^{-1})^{1/2}$, scaling the square root of the Goldreich-Julian corotating polar-cap charge density, and the Langmuir mode growth-rate exponent, is the common parameter for coherent radio emission from both the normal pulsars of Table 2 and the MSP.  It is interesting that such an important conclusion can be drawn from small sets of pulsars which have hitherto appeared to be of little, if any, concern.

\begin{table}
\caption{The Table gives the period $P$ and polar-cap magnetic flux density $B_{12}$ for the MSP drawn from the complete ATNF catalogue as described in the text, having $P < 0.2$ and $BP^{-1} < 2.5\times 10^{10}$ G$s^{-1}$. It is to be recorded that J1641+3627C is in a globular cluster. Each starts with the J-name and columns four and five give, respectively, the observed parameters for pair creation and for the ion-proton model which are inferred to be the threshold values.}
 
\begin{tabular}{lllrc}
\hline
   Pulsar    &      $P$        &    $B_{12}$   &  $B_{12}P^{-2}$ & $(B_{12}P^{-1})^{1/2}$      \\ 
             &   $10^{-3}$ s    &  $10^{-5}$   &                 &                           \\

\hline         
J0514-4002A &   5.0       & 6.0   &  2.4    & 0.11 \\
J0645+5158  &    8.9     &  21   & 2.6    &  0.15  \\
J1101-6424  &   5.1     & 9.7  &  3.7  &  0.14  \\
J1120-3618      &  5.6  & 7.3  &  2.3  &  0.11    \\
J1216-6410     &  3.5  &   7.6   &  6.2  & 0.15    \\
                                                     \\
J1618-4624  &    5.9      &  13.7  &   3.9     & 0.15  \\
J1641+3627C &    3.7      &   6.9 &   5.0    &  0.13 \\
J1801-3210  &   7.5       &  2.3  &  0.41     & 0.18  \\
J1910-5959   & 5.3  &  10.8  &  3.8  &   0.14    \\
J1938+2012  &   2.6       &  4.5 &  6.7  &  0.13 \\
                                                      \\
J2034+3632  &  3.6  &  8.0   &   6.2   &  0.15   \\
J2055+3829  &   2.1       &  4.6  &  10.4   & 0.15  \\
J2212+2450   &  3.9   &   8.9  &  5.8   &   0.17   \\
J2229+2643  &  3.0        &  6.8  &  7.6              & 0.15      \\
J2322-2650  &    3.5      &   4.5  &    3.7            & 0.11     \\          
\hline 
\end{tabular}
\end{table}

The ion-proton model as formulated for pulsars with positive polar-cap corotational charge density (see Jones 2023) has two principal requirements; an accelerated beam of two baryons (protons or ions) having different charge-to-mass ratios and a whole-surface temperature $T_{s}$ high enough to produce photo-electrons as ions accelerate, roughly in the first $10^{6}$ cm of altitude, whose reverse flux creates protons in the surface atmosphere of the neutron star thus automatically giving a source of second baryons in an atmosphere of ions.  The reverse electrons also partially screen the acceleration potential at low altitudes, an important function because the longitudinal-mode growth rate naturally decreases at higher particle Lorentz factors.  It is a self-adjusting process that allows rapid growth of the unstable Langmuir mode. We refer to Jones (2020) for examples. However, the values of $T_{s}$ are not known for either set of pulsars and introduce the question of how the ion-proton conditions can be satisfied.

\begin{table}
\caption{The Table contains all those pulsars in the ATNF catalogue that satisfy the conditions $P >0.2 $ s and $BP^{-1}< 1.0\times 10^{11}$ G$s^{-1}$.   The fourth and fifth columns give, respectively, the observed parameter for pair creation and the observed ion-proton model parameter.}

\begin{tabular}{llcrc}
\hline
pulsar & $P$ &  $B_{12}$  &  $B_{12}P^{-2}$  & $(B_{12}P^{-1})^{1/2}$ \\
        &     s         &          &         &        \\    
\hline
                                                      \\
J0038-2501      &  0.26  &  0.014  &   0.21  &  0.23   \\
J0211+4235 & 0.35  &  0.020  &  0.18 &   0.24   \\
J0815+4611    & 0.43  &  0.042  &  0.23  &    0.31  \\
J0919-6040   &  1.22  &  0.112   &  0.08   &  0.30    \\
J1232-4742    &  1.87  & 0.164  &  0.05   &  0.30   \\
                                                    \\
J1320-3512  &  0.46  &  0.030   &    0.14   & 0.25  \\
J1333-4449  &  0.35   &  0.014  &  0.11  &  0.20   \\
J1514-5316  &  0.30   &  0.022  &  0.24  &  0.27  \\
J1611-5847  &  0.35  &  0.027 &  0.22  &  0.28  \\
J1700-3919   & 0.56  &  0.054  &   0.17 &    0.31   \\
                                                            \\
J1745-2758 & 0.49  &  0.047  &  0.20  & 0.31    \\
J1805-2447 &0.66  &  0.063  & 0.14   &  0.31    \\
J1816-5643 & 0.22 &  0.021  &  0.43   &  0.31   \\
J1904+0056 & 0.44  &  0.042  &  0.22   &  0.31  \\
J1930+1408& 0.43  &  0.029   &  0.16   &  0.26   \\
                                                       \\
J1954+2923 & 0.44  &  0.027  &  0.15    &  0.25  \\
J2150+3427  &  0.65  &  0.049  & 0.12  &  0.27  \\
J2208+4610 &  0.64  &  0.042  &  0.11 &    0.26  \\

\hline
\end{tabular}
\end{table}

The temperature of the whole neutron-star surface that is visible in the frame of the accelerated ion is generally not well known. (Photons from the polar cap do not have a favourable Lorentz transformation to the rest frame of the accelerated ion.)
A survey of theoretical work in this field has been published by Beznogov, Potekhin  and Yakovlev (2021), also a catalogue of thermally emitting isolated neutron stars Potekhin et al (2020). Relatively high surface temperatures have been observed in a small number of pulsars considered to be old and a large number of heating mechanisms have been proposed and studied and are cited in the paper of Gonzalez and Reisenegger (2010). More recent papers (Abramkin et al 2021; Abramkin et al 2022; Schwope et al 2022) have found observational evidence for some kind of internal heating.

In the ion-proton model, the maximum acceleration potential above the polar cap is derived from the Lense-Thirring effect (Beskin 1990; Muslimov and Tsygan 1992; Harding and Muslimov 2001) and is nearly an order of magnitude smaller than that assumed by Philippov, Timokhin and Spitkovsky in the case of negative polar-cap corotational charge density. It is approximately $1250 B_{12}P^{-2}$ GeV. Thus for the first pulsar in Table 2 it is only 259 GeV. In such cases, this is so small that its screening close to the neutron-star polar-cap surface may not be necessary for accelerated ions of two different charge-to-mass ratios rather than an ion-proton combination.

The second major unknown is the composition of the thin atmosphere of the polar cap. A canonical value of atomic number $Z = 26$ has been assumed in previous ion-proton work but there are other possibilities that have been considered, including $Z = 6$. It is certainly true that normal pulsars and the MSP have different histories and may differ in this respect and it is possible that for common ion atomic number, different degrees of ionization at the start of acceleration would satisfy the instability condition of two different charge-to-mass ratios.  We should also mention that, as can be seen in equation (1), the Langmuir mode growth rate exponent also has some dependence on the masses and atomic numbers of the ion(s) concerned. A further unknown relevant to the polar-cap atmosphere is the probable existence of reverse current sheets close to the division between open and closed sectors, impacting on the neutron-star surface and affecting the nature of its surface ions.  This has not been considered in our work.

\section{Conclusions}

But our conclusion must be that the statement of Karastergiou et al is reinforced by comparison of Tables 1 and 2 and further, that it is not possible to conclude other than that the physical process in emission does not involve electron-positron pair production. The approach of this Letter simply uses the observed facts as they stand and raises further questions. The ion-proton model requires a sense of rotation that has a positive polar-cap corotational charge density. If, in any such neutron star, pair creation were possible in principle, its non-appearance would merely reflect the fact that the more efficient screening mechanism of the ion-proton model prevails and is stable. The polarity of observed radio pulsars is an important piece of information in regard to their magnetospheric structure, including regions beyond the light cylinder. Also, the pulsars of both Tables presumably have counterparts in the galaxy with the opposite sign. An electron flux from the polar cap in these cases, unable to produce pairs, does not have any obviously observable emission in the electromagnetic spectrum. These neutron stars, early after formation, are likely to be observable in some part of the electromagnetic spectrum owing to the high potential differences produced by rotation, but our conclusion is that they must quite early in their life-time become unobservable. The absence of pair creation in radio-loud pulsars also has implications for axion searches owing to the lack of high fluctuating electric fields such as are present in the model of Philippov, Timokhin and Spitkovsky (2020). 

\section{Data availability}

The data underlying this work will be shared on reasonable request to the corresponding author.

\section{Acknowledgment}

It is a pleasure to thank the anonymous referee for a supportive review which also much improved the presentation of this work.

\bsp

\label{lastpage}

\end{document}